\documentclass[10pt,conference]{IEEEtran}
\usepackage{amsmath}
\usepackage{amssymb,palatino,theorem}
\usepackage{graphicx}

\def\xv     {\mbox{\boldmath $x$}}
\def\yv     {\mbox{\boldmath $y$}}
\def\nv     {\mbox{\boldmath $n$}}
\def\Hm     {\mbox{\boldmath $H$}}
\def\Xm     {\mbox{\boldmath $X$}}
\def\Ym     {\mbox{\boldmath $Y$}}
\def\Zm     {\mbox{\boldmath $Z$}}
\def\Nm     {\mbox{\boldmath $N$}}
\def\Pm     {\mbox{\boldmath $P$}}
\def\Km     {\mbox{\boldmath $K$}}

\newtheorem{corollary}{Corollary}

\newtheorem{theorem}{Theorem}

\makeatletter
\def\@begintheorem#1#2{\it \trivlist \item[\hskip \labelsep{\bf #1\
#2.}]} \makeatother

\newcommand{\be}{\begin{equation}}
\newcommand{\ee}{\end{equation}}
\newcommand{\bea}{\begin{eqnarray}}
\newcommand{\eea}{\end{eqnarray}}

\newcommand{\beq}[1]{\begin{equation}\label{#1}}
\newcommand{\eeq}{\end{equation}}

\newcommand{\beqn}[1]{\begin{eqnarray}\label{#1}}
\newcommand{\eeqn}{\end{eqnarray}}

\newcommand{\beaa}{\begin{eqnarray*}}
\newcommand{\eeaa}{\end{eqnarray*}}
\def\E{{\mathbb E}}

\def\SNR    {\mbox{\scriptsize\sf SNR}}

\def\MMSE    {\mbox{\scriptsize\sf MMSE}}

\def \det       {{\rm det}}

\def \nT        {n_\mathrm{\scriptscriptstyle T}}
\def \nR        {n_\mathrm{\scriptscriptstyle R}}

\def \Tr        {\mathrm{Tr}}

\def\non{\nonumber\\}

\def\Idm{{\bf I}}

\def\AbvGT #1#2{\lower2pt\vbox{\baselineskip0pt \lineskip-.5pt%
         \halign{$#1 ##$\cr #2\crcr >\cr}}}

\def\complex{\mathop{\raise .45ex\hbox{${\bf\scriptstyle{|}}$}
      \kern -0.40em {\rm \textstyle{C}}}\nolimits}
\def\hilbert{\mathop{\raise .21ex\hbox{$\bigcirc$}}\kern -1.005em
{\rm\textstyle{H}}} %Hilbert space
\begin{document}

\title{What is the Value of Joint Processing of Pilots and Data in Block-Fading Channels?}

\author{
\authorblockN{Nihar Jindal}
\authorblockA{University of Minnesota\\
Minneapolis, MN 55455\\
Email: nihar@umn.edu}
\and
\authorblockN{Angel Lozano}
\authorblockA{
Universitat Pompeu Fabra\\
Barcelona 08003, Spain\\
Email: angel.lozano@upf.edu
\and
\authorblockN{Thomas L. Marzetta}
\authorblockA{Bell Labs (Alcatel-Lucent) \\
Murray Hill, NJ 07974, USA\\
Email: tlm@research.bell-labs.com}
}}

\maketitle

\begin{abstract}
The spectral efficiency achievable with joint processing of pilot and data symbol observations is compared with that
%spectral efficiency
achievable through the conventional (separate) approach of first estimating the channel on the basis of the
pilot symbols alone, and subsequently detecting the data symbols.  Studied on the basis of a mutual
information lower bound, joint processing is found to provide a non-negligible advantage relative to separate processing, particularly
for fast fading. It is shown that, regardless of the fading rate, only a very small number of pilot symbols (at most one per transmit antenna and per channel coherence interval) should be transmitted if joint processing is allowed.
\end{abstract}

\section{Introduction}
Pilot symbols (a.k.a. training or reference symbols) are an inherent part of virtually every wireless system. Motivated by this prevalence, the spectral
efficiency achievable when coherently detecting data with the assistance of pilots has been the object of much analysis (e.g.,
\cite{medard}--\nocite{Zheng02,hassibi,Furrer07}\cite{amos-shamai}). A large fraction of such work has focused on the spectral efficiency achievable with
Gaussian inputs under the assumption that the fading channel is estimated on the basis of the pilot observations and then, using such estimate as it were
the true channel, the data is detected. Although suboptimal, such separate processing reflects the operating conditions of existing systems.

In this paper, we move beyond this approach and quantify the
advantage of jointly processing pilot and data observations when
Gaussian codebooks are utilized. Since the general mutual
information expression is intractable, we rely
%---as did the aforementioned references---
on lower bounds to the achievable
spectral efficiency. These bounds allow assessing the optimum number
of pilot symbols under such joint processing, and also quantify the
minimum improvement in spectral efficiency that joint processing
brings about relative to separate processing.

Although there has been prior work on
receiver design for joint processing (e.g.,
\cite{Tong04}-\nocite{ZhangLaneman07}\cite{LiCollins05}), to the best of
our knowledge there is not yet a general understanding of
the conditions (in terms of signal-to-noise ratio, fading rate, and
antenna configurations) in which joint processing provides a substantial improvement.
Given that joint processing is more complex than separate processing,
such a quantification appears very useful.

As a starting point, a simple block-fading ergodic channel model is
considered. Section \ref{padel} restricts itself to scalar channels,
from which many of the insights can already be derived. The
generalization to MIMO (multiple-input multiple-output) follows in
Section \ref{foolsday}.
%The power is held constant during
%transmission of both pilot and data symbols, with the possibility of
%pilot power boosting left as a future extension.

\section{SISO}
\label{padel}
\subsection{Channel Model}

Let $H$ represent a discrete-time scalar fading channel.
Under block Rayleigh-fading, the channel is drawn from a zero-mean complex Gaussian distribution at the beginning of each
block and it then remains constant for the $T$ symbols composing the block, where $T$
corresponds to the coherence time/bandwidth. This process is repeated for every block in an IID
(independent identically distributed) fashion.
%and thus the model applies to either time- or frequency-domain coding
%depending on whether $T$ signifies the coherence time or the coherence bandwidth.
A total of $\tau$ pilot symbols are inserted within each block leaving $T-\tau$ symbols available for data.
%\footnote{Under block fading, their precise placement is immaterial. In a continuous-fading setting, they would have to be properly spaced.} The rest, $(T-\tau)$ symbols, contain data.

During the transmission of pilot symbols,
\be
\label{siso1}
\yv_{\sf p} = \sqrt{\SNR} \, H + \nv_{\sf p}
\ee
where the received signal, $\yv_{\sf p}$, and the noise, $\nv_{\sf p}$, are $\tau$-dimensional vectors.
The entries of $\nv_{\sf p}$ are IID zero-mean unit-variance complex Gaussian.
The channel satisfies $\E[|H|^2]=1$ and thus $\SNR$ indicates the average signal-to-noise ratio.
During the transmission of data symbols
\be
\label{siso2}
\yv_{\sf d} = \sqrt{\SNR} \, H \xv + \nv_{\sf d}
\ee
where $\yv_{\sf d}$, $\nv_{\sf d}$, and the transmitted data $\xv$, are all $(T-\tau)$-dimensional. The noise $\nv_{\sf d}$ is independent of $\nv_{\sf p}$ but it abides by
the same distribution.
As argued in the Introduction, the entries of $\xv$ are IID zero-mean unit-variance complex Gaussian.
Each transmitted codeword spans a large number of fading blocks, which endows ergodic quantities with operational meaning.

\subsection{Perfect CSI}

If the receiver is provided with perfect CSI (channel-state information), Gaussian codebooks are capacity-achieving
and the ergodic capacity, in bits/s/Hz, equals
\begin{eqnarray}
C(\SNR) &=& \E \left[ \log_2 \left(1 + \SNR \, |H|^2 \right) \right] \\
&=& e^{1/\SNR} E_1\! \left( \frac{1}{\SNR} \right) \log_2 e
\end{eqnarray}
where $E_k(\cdot)$ is the exponential integral of order $k$.
For compactness, $C(\SNR)$ is often abbreviated as $C$.

\subsection{Separated Processing of Pilots and Data}
\label{palomar}

If the receiver uses the pilot observations, $\yv_{\sf p}$, to first
produce an MMSE estimate of the channel, $\hat{H}$, and then performs
nearest-neighbor decoding while treating $\hat{H}$ as if it were $H$,
the maximum spectral efficiency is \cite{amos-shamai}
\begin{eqnarray} \label{IS}
I_{\sf S} = \max_{\tau: 1 \leq \tau <T} \left\{ \left( 1 - \frac{\tau}{T} \right) C \left( \SNR_{\sf eff} \right) \right\}
\end{eqnarray}
with
\be \label{tertulia}
\SNR_{\sf eff} = \frac{\SNR \, (1-\MMSE)}{1+\SNR \cdot \MMSE}
\ee
and
$\MMSE = \E [ | H - \hat{H} |^2 ] = 1/(1 + \SNR \, \tau )$.
The maximization in (\ref{IS}) must be computed numerically as no closed form exists.

\subsection{Spectral Efficiency Lower Bounds for Joint Processing}

In the general case, the receiver decodes the data based upon $\yv_{\sf p}$ and $\yv_{\sf d}$
without any constraints on how these observations are used.  The per-symbol mutual information $ I(\xv ; \yv_{\sf p}, \yv_{\sf d})/T$ is the maximum achievable spectral efficiency
and is achieved by a maximum-likelihood decoder based on the true channel description $p(\yv_{\sf p}, \yv_{\sf d} | \xv)$.
Since the expression for this mutual information is intractable, we instead utilize the following lower bound.

\begin{theorem}
The ergodic spectral efficiency in bits/s/Hz when $\tau$ pilot symbols and $(T - \tau)$ complex Gaussian data symbols are transmitted on every fading block and jointly processed
at the receiver satisfies
\begin{eqnarray}
\frac{1}{T} \, I(\xv ; \yv_{\sf p}, \yv_{\sf d}) \geq I_{\sf J_1} \geq I_{\sf J_2}
\end{eqnarray}
where
\be
I_{\sf J_1} = \left( 1 - \frac{\tau}{T} \right) C - \frac{\log_2 e}{T} \, e^{\tau+1/\SNR}  \sum_{k=1}^{T-\tau} E_k \! \left( \tau + \frac{1}{\SNR}  \right)
\label{etoo}
\ee
and
\be
I_{\sf J_2} = \left( 1 - \frac{\tau}{T} \right) C - \frac{1}{T} \log_2 \left( \frac{1 + \SNR \, T}{1 + \SNR \, \tau} \right).
\label{noucamp}
\ee

{\bf Proof:} See Appendix A.
\end{theorem}

The bound $I_{\sf J_1}$ (or, more precisely, its MIMO form given in Section \ref{foolsday}) was first derived in \cite{Furrer07}. However, it was not given as in (\ref{etoo}) but rather left as an expectation over
the distribution of $\xv$. As shown in the Appendix, where we provide an alternative derivation, this
expectation can be expressed in closed form using the results of \cite{shinlee}.

%The bound $I_{\sf J_1}$ (or, more precisely, its MIMO form given in Section \ref{foolsday}) was derived in \cite{Furrer07, Marzetta08}.
%In both cases, the bound was derived in a different manner and was not given as in (\ref{etoo}) but rather left as an expectation over
%the distribution of $\xv$. As shown in the Appendix, this
%expectation can be expressed in closed form using the results of \cite{shinlee}.

When no pilots are transmitted ($\tau=0$),  $I_{\sf J_1}$ reduces to the bound given for data-only transmission in \cite{Godavarti03}.

\subsection{Optimization of Number of Pilot Symbols}
\label{iniestaisback}

An initial assessment of the optimum number of pilot symbols can be made on the basis of $I_{\sf J_2}$, whose maximization w.r.t. $\tau$ reduces to maximizing
the concave function $\log_2(1+\SNR \, \tau) -  \tau \, C$. By relaxing $\tau$ to a continuous value, the optimum number of pilots is
\be
\tau^{\star} = \frac{\log_2 e}{C} - \frac{1}{\SNR}
\ee
which satisfies $0 \leq \tau^\star \leq 1$. This points to $\tau^{\star}$ being, when restricted to integers,
either $0$ or $1$. Furthermore,
$C < \log_2(1+\SNR)$ (by Jensen's) implying $\tau^{\star}=1$.
%that $\tau=1$ is the maximizer.

In order to sharpen the above assessment, we turn to the tighter $I_{\sf J_1}$ and consider the low- and high-power regimes separately.
In the low-power regime, using
\be
C = \log_2(e) \left( \SNR - \SNR^2  \right) + \mathcal{O}(\SNR^3)
\ee
and
\be
e^{\tau+1/\SNR} E_k \! \left( \tau + \frac{1}{\SNR} \right) = \SNR - (k+\tau) \, \SNR^2 + \mathcal{O}(\SNR^3)
\ee
it is found that maximizing $I_{\sf J_1}$ to second order entails maximizing the concave function $(T-\tau)(T+\tau-1)$. Thus,
the optimum is again either $\tau=0$ or $\tau=1$. While both values yield the same $I_{\sf J_1}$ to second order, an exact computation
of (\ref{etoo}) reveals that $\tau^\star=1$ for $\SNR \rightarrow 0$.

In the high-power regime, using
\bea \label{xmas}
\!\! e^{1/\SNR} E_1(1/\SNR) \!\! & \!\! = \!\! & \!\! \log_2 \SNR - \gamma \log_2 e + \mathcal{O} \left(\frac{1}{\SNR} \right) \;\; \\
\!\! e^{1/\SNR} E_k(1/\SNR)  \!\! & \!\! = \!\! & \!\! \frac{1}{k-1} + \mathcal{O} \left(\frac{1}{\SNR} \right), \quad k>1, \;\;
\eea
where $\gamma=0.5772...$ is the Euler-Mascheroni constant,
%and recalling that $C &=& e^{1/\SNR} E_1(1/\SNR)$,
%\bea
%\label{xmas}
%C &=& e^{1/\SNR} E_1(1/\SNR) \\
%&=& \log_2 \SNR - \gamma \log_2 e + \mathcal{O} \left( \frac{1}{\SNR} \right)
%\eea
it is found that
%, for $\tau=0$,
\begin{eqnarray}
I_{\sf J_1}|_{\tau=0} &=& \frac{T-1}{T} \, C - \frac{\log_2 e}{T} \sum_{k=1}^{T-1} \frac{1}{k} \\
%\ee
%while, for $\tau=1$,
%\be
I_{\sf J_1}|_{\tau=1} &=& \frac{T-1}{T} \, C - \frac{\log_2 e}{T} \sum_{k=1}^{T-1} e \cdot E_k(1) .
\end{eqnarray}
Since $e \cdot E_k(1) < 1/k$ strictly, $\tau = 1$ is preferrable over $\tau=0$ for $\SNR \rightarrow \infty$. (For $\tau \geq 2$, $I_{\sf J_1}$ falls rapidly.)

Altogether, the optimum number of pilots is $\tau^\star=1$ in both the low- and high-power regimes.
Setting $\tau=0$ results in a slight loss (quantified in Section \ref{NJ}), whereas $\tau \geq 2$ is decidedly suboptimal at
moderate/high SNR.

Extrapolating this result to more realistic continuous-fading channels (i.e., the channel varies from symbol-to-symbol according
to a random process), we can infer that, with joint processing, it is desirable to have at most roughly one pilot symbol per coherence interval.

\subsection{Comparison with Separate Processing of Pilots and Data}

\begin{figure}
  \begin{center}
 \includegraphics[width=3.2in]{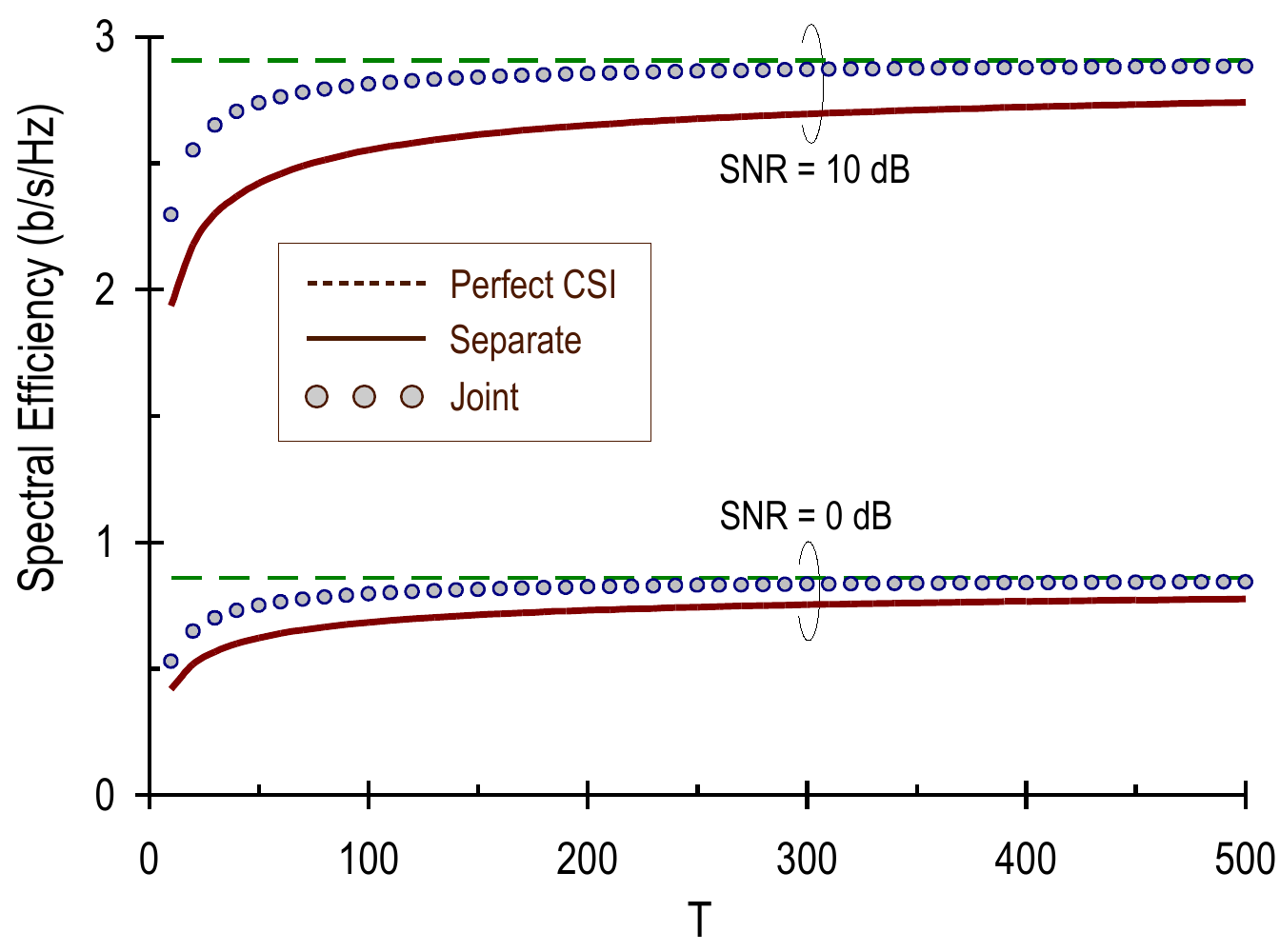}
      \end{center}
  \caption{Spectral Efficiency vs. $T$ for a SISO channel at $\SNR=0$ dB and $\SNR=10$ dB. The curves correspond to $C$, $I_{\sf S}$ and $I_{\sf J_1}$
  (with $\tau=1$).}
  \label{fig-Tscaling}
\end{figure}

The value of joint processing is illustrated by examining how the spectral efficiency converges to the perfect-CSI capacity
as the blocklength $T$ increases. From (\ref{noucamp}), the difference between $C$ and $I_{\sf J_2}$ is
\bea
C - I_{\sf J_2} \!\! & \!\! = \!\! & \!\!  \frac{\tau}{T} \, C + \frac{1}{T} \log_2 \left( \frac{1 + \SNR \, T }{1 + \SNR \, \tau} \right) \\
 \!\! & \!\! = \!\! & \!\!  \mathcal{O} \left( \frac{ \log_2 T}{T} \right)
\eea
for any fixed value of $\tau$. On the other hand, the difference between $C$ and
the spectral efficiency achievable with separate processing, $I_{\sf S}$,
%(with either an optimized number of pilot symbols or a single power-boosted pilot)
vanishes only as $\mathcal{O} ( 1/\sqrt{T} )$ \cite{JindalLozano09}.
This contrast is evidenced in Fig. \ref{fig-Tscaling}.
%, where the spectral efficiency with joint processing converges rapidly to $C$
%while the convergence is much slower with separate channel estimation and data detection.

With joint processing, as $T$ grows the spectral efficiency converges to $C$ even though $\tau$ is fixed because the
(possibly implicit) channel estimation process can take advantage of the data symbols.
On the other hand, if $\tau$ were kept fixed the spectral efficiency of the separate approach would not converge to $C$;
$I_{\sf S}$ converges to $C$ only because $\tau$ is properly increased, as per (\ref{IS}), with $T$.
%because the channel estimation quality is completely determined by $\tau$.
%the number of pilot symbols
%must be appropriately increased with $T$ in order for $I_{\sf S}$ to converge to $C$.
%Indeed, if $\tau$ is kept fixed and the separated approach is used, the spectral efficiency does not converge to $C$
%because the channel estimation quality is completely determined by $\tau$.

\subsection{High-Power Behavior}
\label{NJ}

Further insight is obtained
by studying the high-power behavior of the various bounds. At high $\SNR$, and
for $\tau = 1$, the lower bounds converge absolutely to
% (in an absolute sense) to
\begin{eqnarray}
I_{\sf J_1} \!\! & \!\! \rightarrow \!\! & \!\! \frac{T-1}{T} \left( C - \frac{e \log_2(e) \sum_{k=1}^{T-1} E_k (1)}{T-1 } \right) \\
I_{\sf J_2} \!\! & \!\! \rightarrow \!\! & \!\! \frac{T-1}{T} \left( C - \frac{ \log_2 T}{T-1 } \right),
\end{eqnarray}
while, with separate processing \cite{hassibi},
\be
I_{\sf S} \rightarrow \frac{T-1}{T} \left( C - 1 \right).
\ee
%I_{\sf S_B} \!\! & \!\! \rightarrow \!\! & \!\! \frac{T-1}{T} \left( C - 2 \log_2 \! \left( \sqrt{1 - \frac{1}{T}} + \sqrt{ \frac{1}{T}} \right) \right) .
%where the second expression comes from [Equation 30, Hassibi-Hochwald].

All the above quantities have the same pre-log factor, $(T-1)/T$, and thus the difference between the terms inside the brackets directly gives the power penalty relative to the perfect-CSI capacity,
i.e., the horizontal shift in a plot of spectral efficiency vs. $\SNR$ (dB).
When the information units are bits, this horizontal shift is in $3$-dB units \cite{poweroffset}.

The asymptotic difference between $I_{\sf J_1}$ and $I_{\sf J_2}$ is
\be
\frac{1}{T-1} \left( \log_{2} T -  e \log_{2}(e) \sum_{k=1}^{T-1} E_k (1)  \right)  ,
\ee
in $3$-dB units.
This quantity decreases with $T$ and is minute even for small values of $T$ (e.g., $0.02$ dB for $T=10$) and thus, at high $\SNR$,  we can consider
%the high-power behavior via
the simpler $I_{\sf J_2}$ with only a negligible loss in accuracy.

\begin{figure}
    \begin{center}
   \includegraphics[width=3.2in]{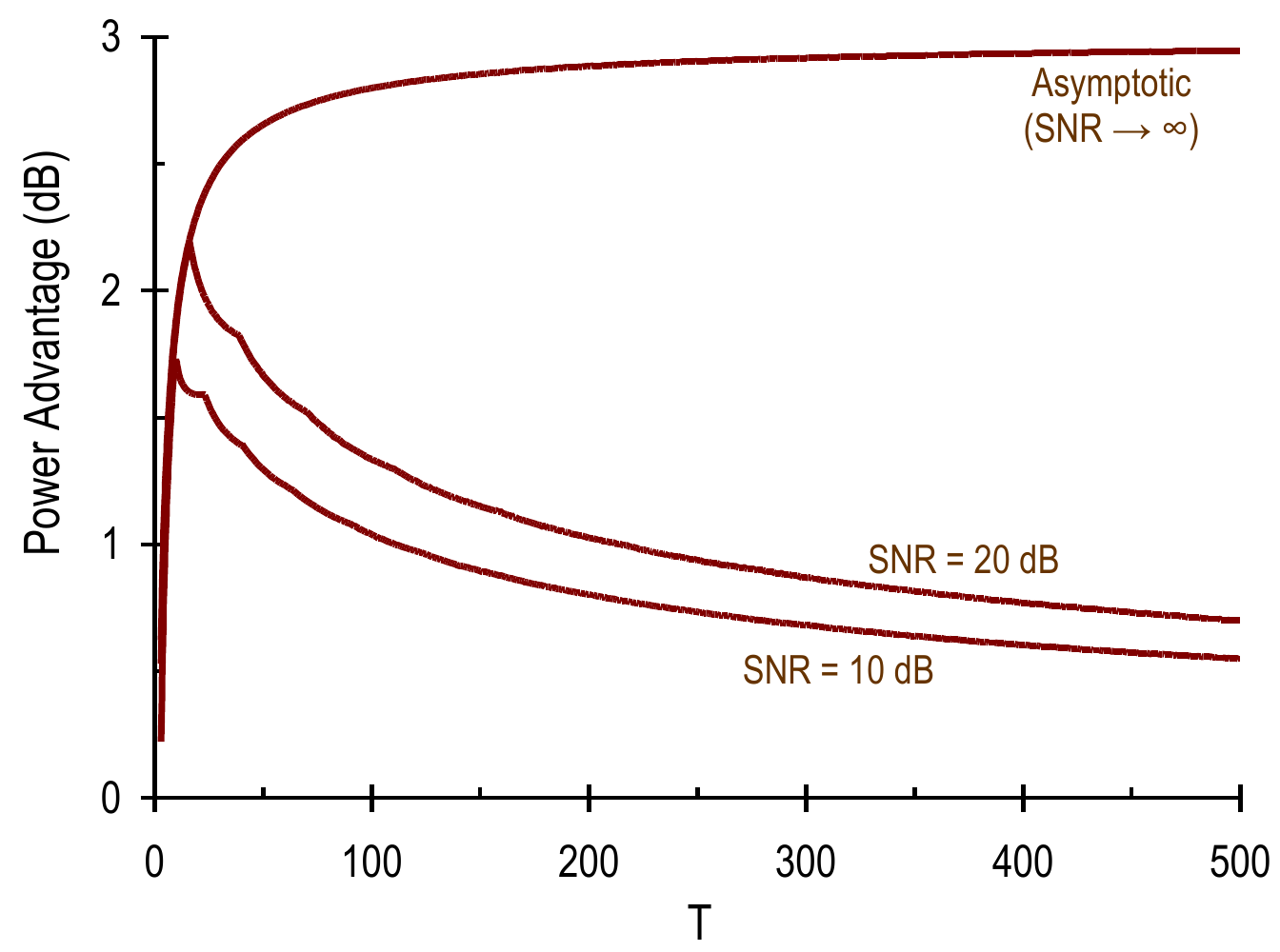}
        \end{center}
    \caption{Power advantage of joint relative to separate processing asymptotically ($\SNR \rightarrow \infty$) and at $\SNR=10$ dB and $\SNR=20$ dB.}
    \label{fig-power}
\end{figure}

Based on $I_{\sf J_2}$ then, the asymptotic power advantage of joint processing relative to separate is
\be
\label{cafe1}
1 - \frac{ \log_{2} T}{T-1}
\ee
in $3$-dB units.
In Fig. \ref{fig-power}, this quantity is plotted versus $T$, along with the numerically computed advantage at $\SNR=10$ dB and $\SNR=20$ dB.
(The difference between the respective curves indicates that the convergence of
$I_{\sf S}$ to its asymptote occurs ever more slowly as $T$ grows.)

Using $I_{\sf J_2}$ and (\ref{xmas}), it is also straightforward to compute the high-power advantage of
transmitting one pilot symbol ($\tau=1$) rather than none ($\tau=0$) as
\be
\frac{\gamma \log_{2} e}{T}
\ee
in $3$-dB units.
For short blocks the single pilot is useful, but for larger blocklengths it makes little difference.

Finally, we can also quantify the distance to the true capacity of the block-fading channel.
In \cite{hochwald}, such capacity (indicated by $\mathcal{C}$ to distinguish it from $C$, the capacity with perfect CSI)
is shown to converge, for $\SNR \rightarrow \infty$, to
\be
\mathcal{C} \rightarrow \frac{T-1}{T} \left( C - \frac{1}{T-1} \log_2 \! \left( \frac{ e^{T-1} (T-1)! }{T^{T-1}} \right) \right) .
\ee
Using Stirling's approximation,
%\be
%\log_2 \left( \frac{ e^{T-1} (T-1)! }{T^{T-1}} \right) \approx \frac{1}{2} \log_2 (T-1)
%\ee
\be
\mathcal{C} \approx  \frac{T-1}{T} \left( C - \frac{1}{2}\frac{ \log_2 T}{T-1 } \right)
\ee
for large $\SNR$, coinciding with the high-$\SNR$ expansion of $I_{\sf J_2}$ save for the
factor $1/2$. This indicates that the spectral efficiency with joint processing
scales with the blocklength $T$ in the same manner as the true capacity in the
high-power regime. Furthermore, the power offset between $I_{\sf J_2}$ and the true capacity is only
(approximately)
\be
\frac{1}{2} \frac{\log_2 T }{T-1}
\ee
in $3$-dB units. This evaluates, for instance, to $0.55$ dB and $0.1$ dB for $T=10$ and $T=100$, respectively.

\section{Generalization to MIMO}
\label{foolsday}

\subsection{Channel Model}

With $\nT$ transmit and $\nR$ receive antennas, the SISO input-output relationships in (\ref{siso1}) and (\ref{siso2}) become
\bea
\Ym_{\sf p}  \!\! & \!\! = \!\! & \!\!  \sqrt{\frac{\SNR}{\nT}} \Hm \Pm + \Nm_{\sf p} \\
\Ym_{\sf d}  \!\! & \!\! = \!\! & \!\!  \sqrt{\frac{\SNR}{\nT}} \Hm \Xm + \Nm_{\sf d}
\eea
where $\Hm$, $\Pm$, $\Xm$, $\Nm_{\sf p}$ and $\Nm_{\sf d}$ are, respectively, $\nR \times \nT$, $\nT \times \tau$, $\nT \times (T-\tau)$, $\nR \times \tau$ and $\nR \times (T-\tau)$.
Matrices $\Hm$, $\Xm$, $\Nm_{\sf p}$ and $\Nm_{\sf d}$ have IID zero-mean unit-variance complex Gaussian entries while $\Pm$ must satisfy
power constraint $\Tr \{ {\bf P} {\bf P}^{\dagger} \} \leq \nT \tau$.

%\be
%\label{santesteve}
%\frac{1}{\tau} \Pm^\dagger \Pm = \Idm .
%\ee

\subsection{Perfect CSI}

For notational convenience, define $C_{t,r}$ as the function
\be
C_{t,r}(\rho) = \E \left[ \log_2 \det \left( \Idm + \frac{\rho}{t} \, \Zm \Zm^\dagger \right) \right]
\label{rain}
\ee
where $\Zm$ is an $r \times t$ matrix with IID zero-mean unit-variance complex Gaussian entries.
%A closed form for $C_{t,r}(\cdot)$ can be found in \cite{shinlee}.
%as
%\bea
%C_{t,r}(\rho) \!\! & \!\! = \!\! & \!\! \log_2(e) \; e^{t/\rho} \sum_{i=0}^{\alpha-1} \sum_{j=0}^i \sum_{\ell=0}^{2j}
%\left[
%\left( \begin{array}{c}
%                2i-2j \\
%                i-j
%              \end{array} \right)
%\right. \non
%&& \cdot \left( \begin{array}{c}
%                2j+2\alpha-2\beta \\
%                2j-\ell
%              \end{array} \right)
%\frac{(-1)^{\ell} \, (2j)! \, (\alpha-\beta+\ell)!}{2^{2i-\ell} \, j! \, \ell! \, (\alpha-\beta+j)!}
%\non
%&& \left. \cdot \sum_{k=0}^{\alpha-\beta+\ell} E_{k+1} \! \left( \frac{t}{\rho} \right)
%\right]
%\eea
%where $\alpha=\min(t,r)$ and $\beta=\max(t,r)$.
The MIMO perfect-CSI capacity with $\nT$ transmit and $\nR$ receive antennas at $\SNR$ equals $C_{\nT,\nR}(\SNR)$.

\subsection{Separated Processing of Pilots and Data}

The SISO expressions for $I_{\sf S}$ in Section \ref{palomar} apply verbatim with $T$, $\tau$, and $C(\cdot)$ replaced, respectively, by
$T/\nT$, $\bar{\tau}=\tau/\nT$, and $C_{\nT,\nR}(\cdot)$.

\subsection{Spectral Efficiency Lower Bounds for Joint Processing}

In the MIMO case, we allow for the possibility of either no pilot symbols ($\tau=0$) or of at least one pilot symbol per
antenna ($\tau\geq\nT$).
%\footnote{If $0<\tau<\nT$, only a subset of transmit antennas can be effectively trained.}

\begin{theorem}
Let $\tau=0$ or $\tau \geq \nT$.
The ergodic spectral efficiency in bits/s/Hz when $\tau$ pilot symbols and $(T - \tau)$ complex Gaussian data symbols are transmitted on every fading block and jointly processed
at the receiver satisfies
\be
\frac{1}{T} \, I(\Xm ; \Ym_{\sf p}, \Ym_{\sf d}) \geq I_{\sf J_1} \geq I_{\sf J_2}
\ee
where
\be
I_{\sf J_1} = \left( 1 - \frac{\tau}{T} \right) C_{\nT,\nR}(\SNR) - \frac{\nR}{T} \, C_{\nT,T-\tau}\left( \frac{\SNR}{1+\frac{\SNR}{\nT} \tau}  \right)
\ee
and
\be
I_{\sf J_2} = \left( 1 - \frac{\tau}{T} \right) C_{\nT,\nR}(\SNR) - \frac{\nT \nR}{T} \log_2 \! \left( \frac{1 + \SNR \, \frac{T}{\nT}}{1 + \SNR \, \frac{\tau}{\nT}} \right)
\ee

{\bf Proof:} See Appendix B.
\end{theorem}

As a by-product of the proof, we show that $I_{\sf J_1}$ is maximized when the pilot matrix $\Pm$ satisfies
\be
\Pm \Pm^\dagger = \tau \Idm
\ee
which coincides with the optimality condition derived in \cite{hassibi} for the case of separate processing.

Henceforth, we shall focus on the case $\nT=\nR$.

\begin{corollary}
\label{wendell}
If $\nT = \nR = n$, then
\be
\frac{I_{\sf J_2} }{n} = \left( 1 - \frac{\tau / n}{T / n} \right) \frac{C_{n,n}(\SNR)}{n} - \frac{1}{T/n} \log_2 \! \left( \frac{1 + \SNR \, T /n}
{1 + \SNR \, \tau /n } \right)
\ee
which coincides with its SISO counterpart in (\ref{noucamp}) only with an effective fading blocklength of $T/n$, an effective number of
pilot symbols of $\tau/n$, and $C$ replaced by $C_{n,n} / n$.
\end{corollary}

%For $\nT = \nR$, therefore, the SISO insights obtained on the basis of $I_{\sf J_2}$ immediately carry over to MIMO.

\subsection{Optimization of Number of Pilot Symbols}

In the low-power regime, the number of pilot symbols can be optimized on the basis of $I_{\sf J_1}$.
Using
\be
C_{t,r}(\rho) = r \log_2 (e) \left( \rho - \frac{t+r}{2 \, t} \, \rho^2 \right) + \mathcal{O}(\rho^3)
\ee
it is found that maximizing $I_{\sf J_1}$ to second order requires maximizing the concave function $(T-\tau)(T+\tau-\nR)$.
This implies that either $\tau=0$ or $\tau = n$ is optimal, and the two are indistinguishable to second order.

%With $\tau$ relaxed to continuous values, this yields $\tau^{\star} = \nR/2$.
Drawing parallels with its SISO counterpart, the maximization of $I_{\sf J_2}$ w.r.t. to $\tau$ is equivalent to the maximization of
$\log_2 \left(1 + \SNR \, \overline{\tau} \right) - \overline{\tau} \, C_{n,n} /n$
w.r.t  $\overline{\tau} = \tau / n$.  Hence,
\be
\overline{\tau}^\star = \frac{\log_2 e}{C_{n,n}/n} - \frac{1}{\SNR}
\ee
if $\overline{\tau}$ is relaxed to continuous values. This quantity is below unity
whenever $C_{n,n}/n \geq \log_2 e$, which implies that the optimum number of pilots is either
$0$ or $n$.
Since $C_{n,n}/n \leq \log_2(1 + \SNR)$, $\tau=n$ is preferred over $\tau=0$.

\subsection{High-Power Behavior}
Because $I_{\sf J_2}$ and $I_{\sf S}$ mirror their SISO counterparts,
%it is easy to verify that
the asymptotic power advantage (in $3$-dB units) of joint relative to separate processing for MIMO is the SISO advantage
for an effective blocklength of $T/n$, i.e.,
\be
1 - \frac{\log_2 (T / n ) }{T /n - 1}
\ee

\section*{Appendix A}

By the chain rule, the mutual information with perfect receiver knowledge of $H$ expands as
$
I(\xv; \yv_{\sf p}, \yv_{\sf d}, H) = I(\xv; \yv_{\sf p}, \yv_{\sf d}) + I(\xv; H | \yv_{\sf p}, \yv_{\sf d}).
$
Thus,
\begin{eqnarray}
I(\xv; \yv_{\sf p}, \yv_{\sf d})  \!\! & \!\! = \!\! & \!\!  I(\xv;  \yv_{\sf p}, \yv_{\sf d}, H) - I(\xv; H | \yv_{\sf p}, \yv_{\sf d}) \\
 \!\! & \!\! = \!\! & \!\!  I(\xv; \yv_{\sf p}, \yv_{\sf d}, H) - h(H | \yv_{\sf p}, \yv_{\sf d}) \non
&& + h(H | \yv_{\sf p}, \yv_{\sf d}, \xv) \\
 \!\! & \!\! \geq \!\! & \!\!  I(\xv;  \yv_{\sf p}, \yv_{\sf d}, H) - h(H | \yv_{\sf p}) \non
&& + h(H | \yv_{\sf p}, \yv_{\sf d}, \xv)
\label{messi1}
\end{eqnarray}
where $h(\cdot)$ denotes differential entropy and (\ref{messi1}) holds because conditioning reduces entropy.
%The bound $I_{\sf J_1}$ equals the right-hand side of (\ref{messi1}), which we elaborate in the following.

%Since the pilot observations at the receiver are of the form $\sqrt{\SNR} \, H + \nv$, the Gaussian-distributed channel $H$ is observed in Gaussian noise with an effective
The signal-to-noise ratio when estimating $H$ on the basis of $\yv_{\sf p}$ is $\SNR \, \tau$.
Thus, $H | \yv_{\sf p}$ is conditionally Gaussian with variance $1/( 1 + \SNR \, \tau )$ and
therefore
\be
h(H | \yv_{\sf p})
%&=& \log_2 \left( \pi e ~ \sigma^2_{H|\yv_{\sf p}} \right) \\
= \log_2(\pi e)  - \log_2 \left(1 + \SNR \, \tau \right)  .
\label{henry1}
\ee
In turn, the signal-to-noise ratio when estimating $H$ on the basis of $(\yv_{\sf p}, \yv_{\sf d})$, conditioned
on  $\xv_{\sf d}$, is $\SNR \, \tau + \SNR \sum_{k=1}^{T-\tau} |x_k|^2$ and thus
\bea
h(H | \yv_{\sf p}, \yv_{\sf d}, \xv)  \!\! & \!\! = \!\! & \!\!  - \E \left[ \log_2 \left( 1 + \SNR \, \tau + \SNR \sum_{k=1}^{T-\tau} |x_k|^2 \right) \right]  \non
&&  + \log_2(\pi e) .
\label{henry2}
\eea
Using $I(\xv; \yv_{\sf p}, \yv_{\sf d}, H) =  (T - \tau) \, C$, plugging (\ref{henry1}) and (\ref{henry2}) into (\ref{messi1}),
and scaling all the terms by $1/T$,
\be
I_{\sf J_1} =  \left( 1 - \frac{\tau}{T} \right) C - \frac{1}{T} \, \E \left[ \log_2 \left( 1+ \frac{\SNR \sum_{k=1}^{T-\tau} |x_k|^2}{1+\SNR \, \tau} \right) \right].
\label{marquet}
\ee
A closed form for the expectation in (\ref{marquet}) is given in \cite{shinlee}, leading directly to (\ref{etoo}).

The subsequent lower bound, $I_{\sf J_2}$, follows from application of Jensen's inequality to (\ref{marquet}).
Since $\E[ |x_k|^2 ]= 1$,
 \be
 \E \left[ \log_2 \! \left(\! 1 + \frac{\SNR \, \sum_{k=1}^{T-\tau} |x_k|^2 }{ 1 + \SNR \, \tau} \right) \right] \leq \log_2 \! \left( \! 1 + \frac{\SNR \, (T-\tau)}{1 +\SNR \, \tau} \right)
 \ee

\section*{Appendix B}

Starting at (\ref{messi1}), we need only compute
$h({\bf H} | \yv_{\sf p})$ and $h({\bf H} | \yv_{\sf p}, \yv_{\sf d}, \xv)$.
Because the $\nR$ antennas are decoupled when conditioned on either $\yv_{\sf p}$
or $(\yv_{\sf p}, \yv_{\sf d}, \xv)$, these terms can be evaluated separately for each receive antenna.
From \cite{hassibi}, the covariances of one row of ${\bf H}$  conditioned on $\yv_{\sf p}$ and on
$(\yv_{\sf p}, \yv_{\sf d}, \xv)$, respectively, are
\begin{eqnarray}
\Km_{{\bf H} | \yv_{\sf p}}  \!\! & \!\! = \!\! & \!\!  \left( {\bf I} + \frac{\SNR}{\nT} {\bf P} {\bf P}^{\dagger} \right)^{-1} \\
\Km_{{\bf H} | \yv_{\sf p}, \yv_{\sf d}, \xv}  \!\! & \!\! = \!\! & \!\!  \left( {\bf I} + \frac{\SNR}{\nT} \left( {\bf P} {\bf P}^{\dagger} + {\bf X} {\bf X}^{\dagger} \right) \right)^{-1}.
\end{eqnarray}
Defining $\Delta = h({\bf H} | \yv_{\sf p}) - h({\bf H | \yv_{\sf p}, \yv_{\sf d}, \xv})$, we have
%\begin{eqnarray}
%h({\bf H} | \yv_{\sf p}) &=& - \log \left( \left( \pi e \right)^{\nT}
%\det \left( {\bf I} + \frac{\SNR}{\nT} {\bf P} {\bf P}^{\dagger} \right) \right) \\
%h({\bf H} | \yv_{\sf p}, \yv_{\sf d}, \xv) &=& - \E \left[ \log \left( \left( \pi e \right)^{\nT}
%\det \left( {\bf I} + \frac{\SNR}{\nT} \left( {\bf P} {\bf P}^{\dagger} + {\bf X} {\bf X}^{\dagger} \right) \right) \right) \right]
%\end{eqnarray}
%Combining the two gives:
\begin{eqnarray}
\!\! \Delta \!\! & \!\! = \!\! & \!\!  \nR \E \left[ \log \det \Km_{{\bf H} | \yv_{\sf p}, \yv_{\sf d}, \xv} \right] - \nR \log \det \Km_{{\bf H} | \yv_{\sf p}} \\
%&=& \E \left[ \log \left(
%\frac{ \det \left( {\bf I} + \frac{\SNR}{\nT} \left( {\bf P} {\bf P}^{\dagger} + {\bf X} {\bf X}^{\dagger} \right) \right) }
%{\det \left( {\bf I} + \frac{\SNR}{\nT} {\bf P} {\bf P}^{\dagger} \right) } \right) \right] \\
 \!\! & \!\! = \!\! & \!\!  \nR \E \left[ \log \det \! \left( {\bf I} + \! \left( {\bf I} + \frac{\SNR}{\nT} {\bf P} {\bf P}^{\dagger} \right)^{-1} \! \frac{\SNR}{\nT} {\bf X} {\bf X}^{\dagger} \right) \! \right] \;\;\;
\label{messi_hat}
\end{eqnarray}
To obtain $I_{\sf J_1}$ we must find the pilot sequence ${\bf P}$ that minimizes (\ref{messi_hat}).
This amounts to choosing the worst-case noise covariance when the input and the channel are both spatially white.
%Because ${\bf P}{\bf P}^{\dagger}$ is the covariance of an additional noise term,
%this is similar to choosing the worst case noise covariance when the input and the channel are both spatially white.
Since the distribution of ${\bf X}$ is rotationally invariant,
%we first notice that the eigenvectors of ${\bf P}$ do not affect (\ref{messi_hat}). Therefore,
we need only consider diagonal forms for ${\bf P}{\bf P}^{\dagger}$.
%Let us first consider the case where $\tau \geq \nT$.
%In this case ${\bf P}{\bf P}^{\dagger}$ can be any diagonal matrix with non-negative entries, subject to the energy constraint
%$\trace({\bf P} {\bf P}^{\dagger}) \leq \nT \tau$.
To show that the best choice is ${\bf P}{\bf P}^{\dagger} = \tau {\bf I}$, we
apply the argument in \cite[Sec. 4.1]{telatar} to the function
in (\ref{messi_hat}), which is convex w.r.t. ${\bf P}{\bf P}^{\dagger}$.
%$\psi({\bf A}) =
%\E \left[ \log \det \left( {\bf I} + \left( {\bf I} + \frac{\SNR}{\nT} {\bf A} \right)^{-1} \frac{\SNR}{\nT} {\bf X} {\bf X}^{\dagger} \right) \right]$
%using the fact that  is convex in ${\bf A}$ to show that the scaled identity is the minimizer.
With ${\bf P}{\bf P}^{\dagger} = \tau {\bf I}$,
\begin{eqnarray}
\Delta
%&=& \E \left[ \log \det \left( {\bf I} + \left( {\bf I} + \frac{\SNR}{\nT} {\bf P} {\bf P}^{\dagger} \right)^{-1} \frac{\SNR}{\nT} {\bf X} {\bf X}^{\dagger} \right) \right] \\
 \!\! & \!\! = \!\! & \!\!  \nR \E \left[ \log \det \left( {\bf I} + \frac{ \frac{\SNR}{\nT}} {1 + \SNR \frac{\tau}{\nT}} {\bf X} {\bf X}^{\dagger} \right) \right]
\label{cruyff} \\
%&=& \E \left[ \log \det \left( {\bf I} + \frac{ \frac{\SNR}{\nT}} {1 + \SNR \frac{\tau}{\nT}} {\bf X}^{\dagger} {\bf X} \right) \right] \\
 \!\! & \!\! = \!\! & \!\!  \nR C_{\nT, T - \tau} \left(\frac{ \SNR} {1 + \SNR \frac{\tau}{\nT}} \right).
\end{eqnarray}
$I_{\sf J_2}$  is reached by applying Jensen's inequality to (\ref{cruyff}).

\end{document}